\documentclass[%
reprint,
showpacs,
amsmath,amssymb,
twocolumn,
]{revtex4-1}
\usepackage{graphicx}
\usepackage{color}

\begin{document}
\title{An analytical study of mismatched complementary media}
\author{Lin Zhu}
\affiliation{Department of Physics and Astronomy, Key Laboratory of Artificial Structures and Quantum Control (Ministry of Education), Shanghai Jiao Tong University, Shanghai 200240, China}
\author{Xudong Luo}
\email{luoxd@sjtu.edu.cn}
\affiliation{Department of Physics and Astronomy,\\ Key Laboratory of Artificial Structures and Quantum Control (Ministry of Education),\\ Shanghai Jiao Tong University, Shanghai 200240, China}
\author{Hongru Ma}
\affiliation{School of Mechanical Engineering; The State Key Laboratory of Metal Matrix Composites; and Key Laboratory of Artificial Structures and Quantum Control (Ministry of Education), Shanghai Jiao Tong University, Shanghai 200240, China}
\begin{abstract}
Complementary media (CM) interacting with arbitrarily situated obstacles are usually less discussed. In this paper, an analytical framework based on multiple scattering theory is established for analyzing such a mismatched case. As examples, CM-based devices, i.e., a superlens and superscatterer, are discussed. From an analysis, the cancellation mechanism of the mismatched CM is studied. In addition, numerical results are provided for illustration. Moreover, further study shows that such cancellation effects might rely on specific conditions. Actually, the conclusions are not restricted to any specific frequencies; they could be extended to many other areas including applications to active cloaking, antennas, and wireless power transfer.
\end{abstract}
\pacs{42.25.Fx, 81.05.Xj, 42.79.-e, 41.20.Jb}
\maketitle
\section{Introduction}
The concepts of complementary media \cite{pendry2003complementary} (CM) have been applied in many studies\cite{yang2008superscatterer,luo2009conceal,wang2011waveguide,li2010experimental,zhang2010illusion}, and they provide a simple and clear geometric interpretation for the propagation of metamaterial \cite{shamonina2007metamaterials,chen2010transformation} controlled electromagnetic (EM) waves. More prosaically, the CM possess the ability of optical ``cancellation,'' e.g., a well-designed negative index material (NIM) could ``cancel'' a matched medium (or vacuum space), and sometimes such a cancellation could produce an image for an interacted object or hide a certain part of an object \cite{yang2008superscatterer,luo2009conceal,lai2009illusion,lai2009complementary}. Well-known devices such as a superlens and superscatterer \cite{yang2008superscatterer,pendry2000perfectlens} can be easily understood with the concept of CM. Usually, in studies on CM-based optical devices, the situated objects should not have an impact on the cancellation, or the CM should be properly arranged in advance to cancel any placed obstacles \cite{lai2009illusion,lai2009complementary}. On the other hand, very few studies have been carried out for the mismatched case, where objects occupy the space or medium that used to be canceled by the prearranged CM. It is commonly believed that when the condition of CM is not satisfied owing to the misplacement of obstacles, the aforementioned optical cancellation ability might be lost \cite{luo2009conceal}. However, our rigorous analysis in this paper gives different results. 

Actually, even for the detection of weak incident EM waves, a relatively strong field could be detected near the surface of the NIM in the scattering field \cite{yang2008superscatterer}. When obstacles are situated near the surface of the NIM, the interpretations that are related to geometrical optics in the long-wavelength limit cannot give an accurate explanation for this situation. Hence, transformation optics and the ordinary concept of CM \cite{pendry2003complementary}, which is derived in the long-wavelength limit, might not be well applicable for revealing the properties of the mismatched case. To obtain an accurate physical picture for such cases, a rigorous analytical analysis is necessary. However, the discussion for such a case is usually avoided in many studies owing to the lack of precise explanation. In fact, it can be encountered in many research areas. For instance, in research on CM-based wireless power transfer \cite{superlensWPT,2015arXiv150802213Z}, obstacles that are located in the nearby area might interfere with the cancellation of the CM and have an impact on the production of an image. 

In this study, to deal with the mismatched CM, an analytical framework based on rigorous multiple scattering theory has been developed. On the basis of the analytical analysis, more physical properties are expected to be revealed. Indeed, our analysis shows that the cancellation ability will still be available for the mismatched CM under certain conditions. The paper is organized as follows. First, to simplify the approach to our analysis, the interpretation of the cancellation properties in mismatched CM is presented with a simple heuristic model based on a superlens \cite{pendry2000perfectlens}. Then, a more applicable model based on a superscatterer \cite{yang2008superscatterer} is provided. In addition to the rigorous analytical analysis, simulations from COMSOL are also provided to illustrate the results. 

\section{Superlens with obstacles}
\begin{figure}
	\centering
	\fbox{\includegraphics[width=\linewidth]{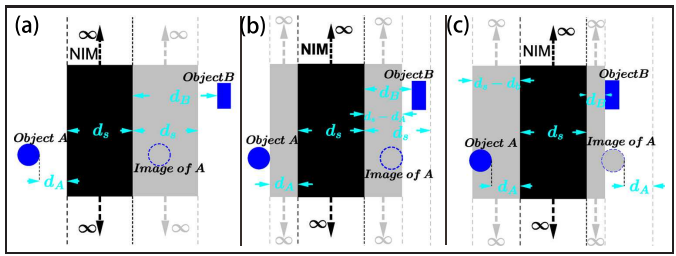}}
	\caption{Schematic of the cancellation strategy in a superlens. (a) The grey domain has not been penetrated by any obstacles; according to the concept of CM, the NIM (domain in black) will cancel the grey domain. (b) If object B is situated near the NIM (i.e., $d_B<d_s$) and the minimum distance between object B and the NIM ($d_B$) is greater than the distance between the image of object A and the NIM (i.e., $d_B>d_s-d_A$), the shape of the grey domain canceled by the black domain is changed according to the position of object A. (c) When object B is located closer to the NIM, i.e., $d_s-d_A>d_B$, the strategy will be changed again, and the shape of the grey domain is determined by the positions of objects A and B.}
	\label{fig:sls}
\end{figure}

To discuss the mismatched CM, a superlens \cite{pendry2000perfectlens} derived from a one-dimensional (1D) transformation could be introduced as a sketchy model at first. Consider the classical superlens shown in Fig. \ref{fig:sls}(a), in which the NIM (domain in black) with a relative permeability and permittivity $\varepsilon=\mu=-1$ will cancel the grey domain (vacuum with $\varepsilon=\mu=1$). According to the interpretation of folded geometry \cite{leonhardt2006general,milton2008solutions}, for an observer on the right side of the grey domain, an image of object A (object situated on the left side of the NIM) can be detected using EM wave detection. When the thickness of the NIM is $d_s$, the distance between object A and its image will be $2 d_s$, as the black domain cancels the grey one. The electric field distributions are provided by simulation in COMSOL and shown in Fig. \ref{fig:sls}(a) and (b).

\begin{figure}
	\centering
	\fbox{\includegraphics[width=\linewidth]{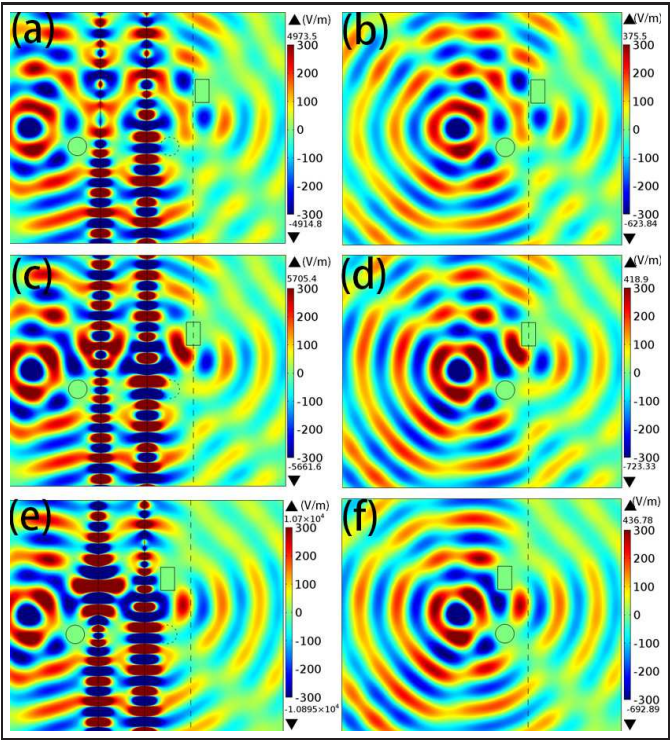}}
	\caption{Simulation results provided by COMSOL for a superlens (with PEC obstacles): When the domain canceled by the slab of the superlens is not penetrated by object B (object B is on the right side of the dashed line), an image of object A can be produced (the distance between object A and its image is $2 d_s$), which is explained by the strategy presented in Fig. \ref{fig:sls}(a). As a result, for an observer on the right side of the canceled domain, the electric field distributions (domain on the right side of the dashed line) in (a) will be identical to those in (b). If object B penetrates the canceled domain, an image of object A can still be detected according to the explanation in Fig. \ref{fig:sls}(b); as a result, the electric field distributions on the right side of the dashed line in (c) are still identical to those in (d). A similar equivalence also exists when object B moves even closer to the NIM slab, as shown in (e) and (f), and can be explained by the strategy in Fig. \ref{fig:sls}(c).}
	\label{fig:sl}
\end{figure}

When the grey domain is penetrated by another object (object B), the cancellation strategy in Fig. \ref{fig:sls}(a) is invalid; however, the simulation shows that the cancellation ability is still available, as shown in Fig. \ref{fig:sl}(c) and (d). In fact, in order to be canceled by the NIM (black domain), the shape of the grey domain should be changed according to the positions of the obstacles. The new cancellation strategy in Fig. \ref{fig:sls}(b) shows that when $d_{s}-d_{A}<d_{B}$ (where $d_A<d_s$), the range of the grey domain will rely on the position of object A ($d_A$). Moreover, when $d_{s}-d_{A}>d_{B}$, as shown in Fig. \ref{fig:sls}(c), the grey domain will be restricted to the positions of both objects A and B ($d_A$ and $d_B$). In other words, when penetrated by object B, the observer will still be able to detect the EM fields scattered by object B and the image of object A, which could not be directly explained from the ordinary cancellation strategy in Fig. \ref{fig:sls}(a). 

Numerical results obtained from the finite element method (FEM) by COMSOL are presented in Fig. \ref{fig:sl}. It should be noted that in the numerical simulation, both objects A and B have perfect electrical conductor (PEC) boundaries, and the results might be different for other types of boundary conditions \cite{2015arXiv150802213Z}, which will be discussed later in this paper. Moreover, it should be emphasized that both transformation optics and the ordinary concept of CM are related to the geometrical optics in the long-wavelength limit; this intuitive explanation works well for the ordinary case, e.g., well-organized CM without the placement of obstacles. However, considering the strong field near the surface of the NIM (shown in Fig. \ref{fig:sl}) in the scattering field \cite{yang2008superscatterer}, the scattering properties of obstacles located near it might not be explained very well by these intuitive theories. Indeed, when obstacles occupy the canceled space near the NIM, multiple scattering theory is necessary for investigating their physical properties.
\section{Superscatterer with obstacles}
\begin{figure}
	\centering
	\fbox{\includegraphics[width=\linewidth]{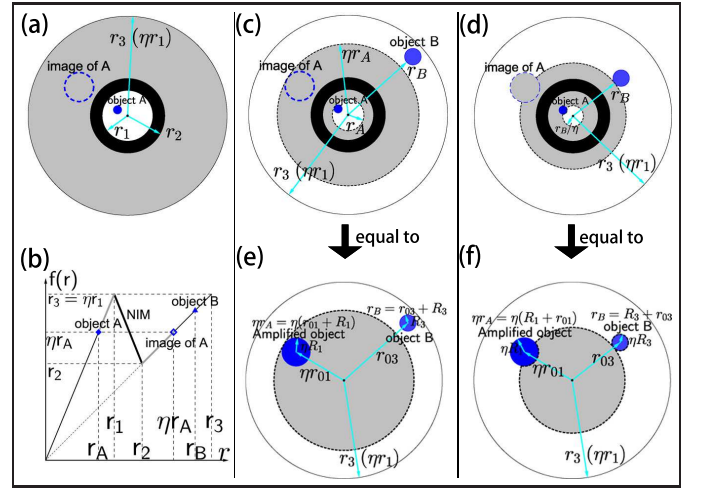}}
	\caption{Schematic of mismatched CM using a superscatterer. The NIM in black (the ss-shell, $r_1<r<r_2$) will cancel the domain painted in grey. Object B and the image of object A do not overlap. (a) Cancellation strategy when no obstacles penetrate the domain $r_2<r<r_3$. (b) and (c) Strategy when the obstacle (denoted as object B with the PEC boundary) penetrates the domain $r_2<r<r_3$, where $r_3>r_B> \eta r_A$. (d) Strategy when $r_B< \eta r_A$. If the cancellation strategy described in (c) is achieved, the scenario in (e) will be the same as the scenario in (c) observed by the viewer in the domain $r>r_3$ for EM wave detection, i.e., an amplified image of object A can be detected. The relationship between (f) and (d) is similar.}
	\label{fig:compleschem}
\end{figure}
To give a clear explanation and to provide a better understanding of the properties of mismatched CM, an analytical framework based on multiple scattering theory should be established. In fact, rather than the superlens discussed above, a two-dimensional (2D) superscatterer \cite{yang2008superscatterer} is more appropriate to show the analytical framework. Since it is a finite circular cylindrical device, it is much easier to give a rigorous analytical analysis here, and transverse electric (TE) or transverse magnetic (TM) wave could be discussed separately in the 2D case. Although the model is relatively simple, the heuristic approach could still provide useful information for three-dimensional (3D) and more practical models. Moreover, the 2D superscatterer has many practical applications in many research areas \cite{luo2009conceal,2015arXiv150802213Z}. 

The analytical analysis for the 2D superscatterer is based on the schematic shown in Fig. \ref{fig:compleschem}. Similar to the analysis in the last section, the black domain is the superscatterer shell (ss-shell) and it is complementary to the grey domain. Further, all obstacles have PEC boundaries. The parameter distributions are derived in cylindrical coordinates, and the relative permeability and permittivity can be deduced as follows:
\begin{equation}
	\left\{ {\begin{array}{*{20}{l}}
		{{\varepsilon _r} = {\mu _r} = \frac{{f(r)}}{r}\frac{1}{{f'(r)}}}, \\ 
		{{\varepsilon _\theta } = {\mu _\theta } = \frac{r}{{f(r)}}f'(r)}, \\
		{{\varepsilon _z} = {\mu _z} = \frac{{f(r)}}{r}f'(r)},
	\end{array}} \right.
	\label{eq:parametersFr}
\end{equation}
with the coordinate transformation
\begin{equation}
	f(r) = \left\{ {\begin{array}{*{20}{l}}
		f_1(r)={\eta r }, \quad\quad\quad\quad {0 < r < {r_1},}\\
		f_2(r)={{r_3} + T(r)},
		\quad {r_1} \le r \le {r_2}, \\
		f_3(r)=r, \quad\quad\quad\quad {r_2} < r < {r_3},
	\end{array}} \right.
	\label{eq:frValue}
\end{equation}
where $\eta=r_3/r_1$, and $T(r)$ could be chosen as any continuous and piecewise differentiable function that makes the domain $r_1<r<r_2$ complementary \cite{pendry2003complementary} to the domain $r_2<r<r_3$ (i.e., $T(r_1)=0, T(r_2)=r_2-r_3 $). For purpose of illustration, we choose $T(r)=\frac{({r} - {r_1})( {r_3} - {r_2})}{{r_1} - {r_2}}$ in this study, as shown in Fig. \ref{fig:compleschem}(b). Thus, the schematic could be divided into three parts; the domain $r_2<r<r_3$ (and $r\ge r_3$) is vacuum, the domain $r_1<r<r_2$ (the shell painted black) is filled with the NIM, and the domain $r<r_1$ is filled with a homogeneous material. From the concepts of CM, as shown in Fig. \ref{fig:compleschem}(a), the domain in black (ss-shell) will cancel the grey domain, and the domain $r<r_1$ (as well as the obstacles inside it) will be amplified in the domain $r<r_3$ (for EM wave detection) with the amplification factor $\eta=r_3/r_1$. As a matter of fact, when the system contains no obstacles, the whole domain $0<r<r_3$ will act as a vacuum observed by the viewer outside. 

It should be noted that if the radius of the ss-shell ($r_1$) becomes infinite and the thickness of the shell and the canceled space are maintained at finite values of $r_3-r_2=d_s$ and $r_2-r_1=d_s$, the superscatterer can be reduced to a slab of the superlens discussed above. Considering $T(r)=\frac{({r} - {r_1})( {r_3} - {r_2})}{{r_1} - {r_2}}$ in Eq. \ref{eq:frValue}, it is easy to deduce ${\lim _{r_1 \to \infty }}(f_2(r)/r) = 1$ for the ss-shell domain, which indicates that the ss-shell becomes a homogeneous slab of a superlens (with a thickness of $d_s$). Moreover, the amplification factor becomes $\eta=r_3/r_1\to 1$. Hence, the parameter distributions derived from Eq. (\ref{eq:parametersFr}) will be ${\varepsilon _r} = {\mu _r}=-1$, ${\varepsilon _\theta } = {\mu _\theta }=-1$, and ${\varepsilon _z} = {\mu _z}=-1$ when ${r_1 \to \infty }$.

Similar to the analysis for the superlens, the obstacles inside and outside the ss-shell are termed as objects A and B, respectively. In addition, $r_A$ is the maximum distance between a point in object A and the center of the ss-shell, $r_B$ is the minimum distance between object B and the center of the ss-shell. As shown in Fig. \ref{fig:compleschem}(a), if no obstacles penetrate the domain $r_2<r<r_3$, a well-matched CM with the black domain that will be complementary to the grey one (domain $r_2<r<r_3$) can be easily satisfied as described by Eq. (\ref{eq:frValue}). As a result, an amplified image of object A with an amplification factor $\eta=r_3/r_1$ could be detected in the domain $r<r_3$. If another object penetrates into the domain $r<r_3$, with $\eta r_A<r_B<r_3$, the aforementioned grey domain in Fig. \ref{fig:compleschem}(a) will be partially occupied, and the ordinary cancellation strategy in Fig. \ref{fig:compleschem}(a) could not be applied. Similar to the discussion of the analysis of the superlens, another strategy could be utilized, as shown in Fig. \ref{fig:compleschem}(c). This strategy seems to be self-adaptive for $r_A$, as the black domain will be complementary to the domains $r_A<r<r_1$ and $r_2<r<\eta r_A$. It should be noted that such a strategy could also explain the amplification effects in Fig. \ref{fig:compleschem}(a) as described in Eqs. (\ref{eq:frValue}) shown in Fig. \ref{fig:compleschem}(b). 

Moreover, when all obstacles have PEC boundaries, object B is situated at $r_B<\eta r_A$, and the third strategy shown in Fig. \ref{fig:compleschem}(d) is more appropriate. It should be remarked that the strategy here is related to both $r_A$ and $r_B$, which implies that the cancellation strategy should rely on the placement of the obstacles, although the parameter distributions in the ss-shell are still the same as above. In fact, the geometric cancellation strategies are just simplified intuitive explanations for the EM wave scattering results; such explanations might not be unique, e.g., the strategy in Fig. \ref{fig:compleschem}(c) (or Fig. \ref{fig:sls}(b)) can also work well in explaining the effects shown in Fig. \ref{fig:compleschem}(a) (or Fig. \ref{fig:sls}(a)). Thus, a rigorous analytical analysis is necessary to understand the physical mechanism of mismatched CM.

\section{Analytical analysis}
To validate the heuristic analysis given in the previous sections, a detailed analytical analysis is presented in this section. Specifically, the obstacles are simplified to circular cylindrical objects, and all obstacles have PEC boundaries. $r_{01}$ (or $r_{03}$) is the distance between the center of object A (or B) and the center of the ss-shell with the angle $\phi_1=\phi_{\overrightarrow{r_{01}}}$ (or $\phi_3=\phi_{\overrightarrow{r_{03}}}$). $R_1$ (or $R_3$) is the radius of object A (or B); thus, we have $r_A=r_{01}+R_1$ and $r_B=r_{03}-R_3$. We consider TE-polarized EM wave detection with the harmonic time dependence $exp(-i\omega t)$. The analytical analysis will be presented as a comparison of the scattering properties of scenario (c) with scenario (e) (or scenario (d) with scenario (f)) in Fig. \ref{fig:compleschem}. More prosaically, our analytical deduction demonstrates that, for detection by the same EM waves (same incident field), scenarios (c) and (e) will give the same scattering EM fields. Such an equivalence can also be found in the comparison between scenarios (d) and (f). The analysis is given for the following four scenarios.

\textbf{For scenarios (c) and (d)}: The general series expansion (analytically deduced from the wave equation \cite{yang2008superscatterer,2015arXiv150802213Z}) for the electric field can be expressed as 
\begin{widetext}
	\begin{footnotesize}
		\begin{equation}
			E_z(r ,\phi ) = \left\{ {\begin{array}{*{20}{l}}
				\sum\limits_{n =  - \infty }^\infty [ {a_n^{(1)}{J_n}({k_0}f_{1}(|\overrightarrow {{r}}|)){e^{in{\phi}}}}+ {b_n^{(1)}{H_n^{(2)}}({k_0}f_{1}(|\overrightarrow {{r}}-\overrightarrow {{r_{01}}}  |)){e^{in{\phi_{(\overrightarrow {{r}}-\overrightarrow {{r_{01}}})}}}}}],\;r<r_1,\\
				\sum\limits_{n =  - \infty }^\infty [ {a_n^{(2)}{J_n}({k_0}f_{2}(|\overrightarrow {{r}}|)){e^{in{\phi}}}}+ {b_n^{(2)}{H_n^{(2)}}({k_0}f_{2}(|\overrightarrow {{r}}|)){e^{in{\phi}}}}],\;r_1<r<r_2,\\
				\sum\limits_{n =  - \infty }^\infty [ {a_n^{(3)}{H_n^{(2)}}({k_0}f_{3}(|\overrightarrow {{r}}|)){e^{in{\phi}}}}+{a_n^{(i)}{J_n}({k_0}|\overrightarrow {{r}}|){e^{in{\phi}}}}+ {b_n^{(3)}{H_n^{(2)}}({k_0}f_{3}(|\overrightarrow {{r}}-\overrightarrow {{r_{03}}})){e^{in{\phi_{(\overrightarrow {{r}}-\overrightarrow {{r_{03}}})}}}}}],\;r>r_2,
			\end{array}} \right.
			\label{eq:Es1}
		\end{equation}
	\end{footnotesize}
\end{widetext}
where $k_0$ is the wave vector in vacuum, and $J_n$ and $H_n^{(2)}$ are the nth-order Bessel function and the nth-order Hankel function of the second kind, respectively. $\alpha=1,2,3$ denotes the domains $r<r_1$, $r_1<r<r_2$, and $r>r_2$, respectively, $a_n^{(\alpha)}$ and $b_n^{(\alpha)}$ are the series expansion coefficients, and $\sum_n{a_n^{(i)}{J_n}({k_0}|\overrightarrow {{r}}|){e^{in{\phi}}}}$ is the incident field, which is chosen to be the same in the following scenarios. Actually, it is sufficient to discuss only the electric field ($z$ component); the corresponding $H_\theta$ components can be derived using the Maxwell equation ${H_\theta } =  - \frac{1}{{i\omega \mu }}\frac{{\partial {E_z}}}{{\partial r}}$. Considering the boundary conditions while imposing Eq. (\ref{eq:frValue}), the coefficients above can be solved from the following equations (where $n=0,\pm 1,\pm 2,\dots$):
\begin{widetext}
	\begin{footnotesize}
		\begin{equation}
			\left\{ {\begin{array}{*{20}{l}}
				{{J_n}({k_0}\eta {R_1})\sum\limits_{m =  - \infty }^\infty  {[a_m^{(1)}{J_{m - n}}({k_0}\eta {r_{01}}){e^{ - i(n - m){\phi _1}}}]}  + b_n^{(1)}{H_n^{(2)}}({k_0}\eta {R_1}) =0},\\
				{{J_n}({k_0}{R_3})\sum\limits_{m =  - \infty }^\infty  {[a_m^{(3)}{H_{m - n}^{(2)}}({k_0}{r_{03}}){e^{ - i(n - m){\phi _3}}}+a_m^{(i)}{J_{m - n}}({k_0}{r_{03}}){e^{ - i(n - m){\phi _3}}}]} + b_n^{(3)}{H_n^{(2)}}({k_0}{R_3}) =0},
				\\
				{\hbox{where}\;\;} {a_m^{(1)} = \sum\limits_{l =  - \infty }^\infty  {[b_l^{(3)}{H_{m - l}^{(2)}}({k_0}{r_{03}}){e^{ - i(m - l){\phi _3}}}]} }+a_m^{(i)} ,\hfill{a_m^{(3)} = \sum\limits_{l =  - \infty }^\infty  {[b_l^{(1)}{J_{m - l}}({k_0}\eta {r_{01}}){e^{ - i(m - l){\phi _1}}}]} }.
			\end{array} } \right.
			\label{eq:meta}
		\end{equation}
	\end{footnotesize}
\end{widetext}
Here, the first and second equations originate from the PEC boundary conditions at the surfaces of the obstacles (objects A and B). The other two are derived from the continuity at the boundaries between the ss-shell and the domains of $r<r_1$ and $r>r_2$. In addition, the translation \cite{chew1995waves} that expresses the wave functions in one coordinate system in terms of those in another coordinate system is applied. The equations can be rewritten in a matrix form as
\begin{footnotesize}
	\begin{eqnarray}
		0=&&[J(\eta {R_1})].[J(\eta {r_{01}},-\phi_1)].[A_{(1)}]  + [H(\eta {R_1})].[B_{(1)}] ,\nonumber\\ 
		0=&&[J({R_3})].([ H({r_{03}},-\phi_3)].[A_{(3)}]+[ J({r_{03}},-\phi_3)].[A_{(i)}])  + \nonumber\\ 
		&&[H({R_3})].[ B_{(3)}],\nonumber\\
		~[A_{(1)}] =&& [ H({r_{03}},\phi_3)].[ B_{(3)}] +[A_{(i)}],\nonumber\\
		~[A_{(3)}] =&& [J(\eta {r_{01}},\phi_1)].[B_{(1)}],
		\label{eq:metaMatrix}
	\end{eqnarray}
\end{footnotesize}
where the matrices $[J(\eta {R_1})]$ and $[J(\eta {r_{01}},-\phi_1)]$ and the vector $[A_{(1)}]$ are defined as follows (and similarly for the others):
\begin{footnotesize}
	\begin{align*}
		&[J(r)] = Diag[ {\begin{array}{*{20}{c}}\cdots, &{{J_{ - n}}({k_0}r)},&\cdots,&{{J_0}({k_0}r)},&\cdots,&{{J_n}({k_0}r)},&\cdots\end{array}}], \\
		&{[J({r}, {\phi})]_{m,n}} = {J_{m - n}}({k_0}{r}){e^{ i(n - m){\phi}}},\\
		&[A_{(1)}] =[{\begin{array}{*{20}{c}}\cdots,&{a_{ - m}^{(1)}},&\cdots,&{a_0^{(1)}},&\cdots,&{a_m^{(1)}},&\cdots 
		\end{array}} ]^T.
	\end{align*}
\end{footnotesize}
Once the scattering coefficients ($[A_{3}]$, $[B_{3}]$) are solved, the EM fields for each scenario can be obtained. However, to discuss the equivalence between scenarios (c) and (e) (or (d) and (f)) for detection of the same EM wave, it is unnecessary to solve for the scattering coefficients; a comparison of the equations that describe the scattering properties of each scenario is sufficient.

On the basis of multiple scattering theory, for scenarios (e) and (f) in Fig. \ref{fig:compleschem}, where $r'_{01}=\eta r_{01}$, $r'_{03}=r_{03}$, $R'_1=\eta R_1$, and $R'_3=R_3$, $E_z(r ,\phi )$ should be expressed as
\begin{footnotesize}
	\begin{eqnarray}
		\label{eq:Esab}
		\sum\limits_{n =  - \infty }^\infty [&& {{b'}_n^{(1)}{H_n^{(2)}}({k_0}(|\overrightarrow {{r}}-\overrightarrow {{r'_{01}}}  |)){e^{in{\phi_{(\overrightarrow {{r}}-\overrightarrow {{r'_{01}}})}}}}}+\\
		&&{{b'}_n^{(3)}{H_n^{(2)}}({k_0}(|\overrightarrow {{r}}-\overrightarrow {{r'_{03}}}  |)){e^{in{\phi_{(\overrightarrow {{r}}-\overrightarrow {{r'_{03}}})}}}}}]+ {a_n^{(i)}{J_n}({k_0}|\overrightarrow {{r}}|){e^{in{\phi}}}}. \nonumber
	\end{eqnarray}
\end{footnotesize}
To compare with scenario (c) (or (d)), we also divide the domain in scenario (e) (or (f)) into two parts: the domains $r<r_{min}$ and $r>r_{min}$, where $r_{min}=Min(r_B, \eta r_A)$. Moreover, we will rewrite Eq. (\ref{eq:Esab}) for these different cases.

\textbf{For scenario (e)}: For $r_B> \eta r_A$ and $r_{min}=\eta r_A$, similar to the above, we can rewrite the electric field as
\begin{widetext}
	\begin{footnotesize}
		\begin{equation}
			E_z(r ,\phi ) = \left\{ {\begin{array}{*{20}{l}}
				\sum\limits_{n =  - \infty }^\infty [ {{a'}_n^{(1)}{J_n}({k_0}(|\overrightarrow {{r}}|)){e^{in{\phi}}}}+ {{b'}_n^{(1)}{H_n^{(2)}}({k_0}(|\overrightarrow {{r}}-\overrightarrow {{{r'}_{01}}}|)){e^{in{\phi_{(\overrightarrow {{r}}-\overrightarrow {{{r'}_{01}}})}}}}}],\;r< r_{min},\\
				\sum\limits_{n =  - \infty }^\infty [ {{a'}_n^{(3)}{H_n^{(2)}}({k_0}(|\overrightarrow {{r}}|)){e^{in{\phi}}}}+{a_n^{(i)}{J_n}({k_0}|\overrightarrow {{r}}|){e^{in{\phi}}}}+ {{b'}_n^{(3)}{H_n^{(2)}}({k_0}(|\overrightarrow {{r}}-\overrightarrow {{r'_{03}}}  |)){e^{in{\phi_{(\overrightarrow {{r}}-\overrightarrow {{r'_{03}}})}}}}}],\;r>r_{min}.
			\end{array}} \right.
			\label{eq:Es2BigB}
		\end{equation}
	\end{footnotesize}
\end{widetext}
Because $r_B>\eta r_A$, object B is in the domain $r>\eta r_A$. Similar to scenarios (c) and (d), the boundary conditions mean that
\begin{footnotesize}
	\begin{eqnarray}
		0=&&[J({R'_1})].[J({r'_{01}},-\phi_1)].[A'_{(1)}]  + [H({R'_1})].[B'_{(1)}],\nonumber\\
		0=&&[J({R'_3})].([H({r'_{03}},-\phi_3)].[A'_{(3)}]+[ J({r'_{03}},-\phi_3)].[A_{(i)}])+\nonumber\\ 
		&&[H({R'_3})].[B'_{(3)}],\nonumber \\
		~[A'_{(1)}] =&& [H({r_{03}},\phi_3)].[B'_{(3)}]+[A_{(i)}] ,\nonumber\\
		~[A'_{(3)}] =&& [J({r'_{01}},\phi_1)].[B'_{(1)}].
		\label{eq:equaMatrixBigB}
	\end{eqnarray}
\end{footnotesize}
We see that the scattering coefficients satisfy exactly the same equations as those in Eq. (\ref{eq:metaMatrix}). This means that $a_n^{(3)}={a'}_n^{(3)}$ and $b_n^{(3)}={b'}_n^{(3)}$ for $n=0,\pm 1,\pm 2,\dots$. Thus, the detected electric fields derived from Eqs. (\ref{eq:Es1}) and (\ref{eq:Es2BigB}) for the domain $r>\eta r_A$ will be the same. In other words, scenario (c) is equivalent to scenario (e) for viewers in the domain $r>Max(\eta r_A,r_2)$.

\textbf{For scenario (f)}: For $r_B<\eta r_A$ and $r_{min}=r_B$, the series expansion for the electric field is different from that in scenario (e), which could be rewritten as follows:
\begin{widetext}
	\begin{footnotesize}
		\begin{equation}
			E_z(r ,\phi ) = \left\{ {\begin{array}{*{20}{l}}
				\sum\limits_{n =  - \infty }^\infty [ {{a'}_n^{(3)}{J_n}({k_0}(|\overrightarrow {{r}}|)){e^{in{\phi}}}}+ {{b'}_n^{(3)}{H_n^{(2)}}({k_0}(|\overrightarrow {{r}}-\overrightarrow {{r_{03}}}  |)){e^{in{\phi_{(\overrightarrow {{r}}-\overrightarrow {{r_{03}}})}}}}}],\;r<r_{min},\\
				\sum\limits_{n =  - \infty }^\infty [ {{a'}_n^{(1)}{H_n^{(2)}}({k_0}(|\overrightarrow {{r}}|)){e^{in{\phi}}}}+{a_n^{(i)}{J_n}({k_0}|\overrightarrow {{r}}|){e^{in{\phi}}}}+ {{b'}_n^{(1)}{H_n^{(2)}}({k_0}(|\overrightarrow {{r}}-\overrightarrow {{{r'}_{01}}}  |)){e^{in{\phi_{(\overrightarrow {{r}}-\overrightarrow {{{r'}_{01}}})}}}}}],\;r> r_{min},
			\end{array}} \right.
			\label{eq:Es2SmallB}
		\end{equation}
	\end{footnotesize}
\end{widetext}
By choosing the virtual interface $r=r_{min}$, the electric field in scenario (f) can be manually divided into two parts, as denoted in Eq. (\ref{eq:Es2SmallB}). However, it should be remarked that $E_z$ in the neighborhood of object B (or A) can always be expressed in the form of the first (or second) equation; thus, its boundary conditions are
\begin{footnotesize}
	\begin{eqnarray}
		0=&&[J({R'_1})].([H({r'_{01}},-\phi_1)].[A'_{(1)}]+[J({r'_{01}},-\phi_1)].[A_{(i)}])+\nonumber\\
		&&[{H}({R'_1})].[B'_{(1)}] ,\nonumber\\ 
		0=&&[J({R'_3})].[ J({r'_{03}},-\phi_3)].[A'_{(3)}]  + [H({R'_3})].[B'_{(3)}],\nonumber\\
		~[A'_{(1)}] =&& [J({r'_{03}},\phi_3)].[ B'_{(3)}],\nonumber\\
		~[A'_{(3)}] =&& [H({r'_{01}},\phi_1)].[B'_{(1)}]+[A_{(i)}].
		\label{eq:equaMatrixSmallB}
	\end{eqnarray}
\end{footnotesize}
It seems that the equations in Eq. (\ref{eq:equaMatrixSmallB}) are different from those in Eq. (\ref{eq:metaMatrix}) and might not give the same solutions. However, we consider the translation relation that expresses wave functions in one coordinate system in terms of the wave functions in another coordinate system \cite{chew1995waves} and remember that the EM wave will remain the same when translated along $\overrightarrow{r'_{0\alpha}}$ and $-\overrightarrow{r'_{0\alpha}}$ ($\alpha=1,3$). Thus, we have the following relationships:
\begin{footnotesize}
	\begin{eqnarray}
		[{H}({{R'}_\alpha})] =&& [J({{R'}_\alpha})].[H({{r'}_{0\alpha}}, - {\phi _\alpha})].[J({{r'}_{0\alpha}},{\phi _\alpha})]\nonumber\\
		=&& [J({{R'}_\alpha})].[J({{r'}_{0\alpha}}, - {\phi _\alpha})].[H({{r'}_{0\alpha}},{\phi _\alpha}].
		\label{eq:tr-aabb}
	\end{eqnarray}
\end{footnotesize}
Similarly, a translation along $-\overrightarrow{r'_{01}}$ then $-\overrightarrow{r'_{03}}$ will be equal to that along $-\overrightarrow{r'_{03}}$ then $-\overrightarrow{r'_{01}}$, which means that
\begin{footnotesize}
	\begin{eqnarray}
		~[J({{r'}_{03}}, - {\phi _3})].&&[H({{r'}_{01}}, - {\phi _1})] =[J({{r'}_{01}}, - {\phi _1})].[H({{r'}_{03}}, - {\phi _3})],\nonumber\\
		~[J({{r'}_{03}}, - {\phi _3})].&&[J,({{r'}_{01}} - {\phi _1})] =[J({{r'}_{01}}, - {\phi _1})].[J({{r'}_{03}}, - {\phi _3})] ,\nonumber\\
		{\hbox{and\;\;}}[V]:=&&{[H({r'_{03}}, - {\phi _3})]^{ - 1}}.[J({r'_{03}}, - {\phi _3})]\nonumber \\
		=&& {[H({r'_{01}}, - {\phi _1})]^{ - 1}}.[J({r'_{01}}, - {\phi _1})].
		\label{eq:tr-ab}
	\end{eqnarray}
\end{footnotesize}
By substituting the related components, Eqs. (\ref{eq:metaMatrix}) and (\ref{eq:equaMatrixSmallB}) can be respectively expressed as
\begin{footnotesize}
	\begin{subequations}
		\begin{equation}
			\left\{ {\begin{array}{*{20}{c}}
				{[H({r_{03}},\phi_3)].[ B_{(3)}] +[A_{(i)}] +[H(\eta {r_{01}},\phi_1)].[B_{(1)}] =0},\\
				{[J(\eta {r_{01}},\phi_1)].[B_{(1)}]+[V].[A_{(i)}] +[J({r_{03}},\phi_3)].[B_{(3)}]=0},
			\end{array}} \right.
			\label{eq:merge-1}
		\end{equation}
		\begin{equation}
			\left\{ {\begin{array}{*{20}{c}}
				{[J({r'_{03}},\phi_3)].[ B'_{(3)}]  +[V].[A_{(i)}] + [J({r'_{01}},\phi_1)].[B'_{(1)}] =0},\\
				{[H({r'_{01}},\phi_1)].[B'_{(1)}]  +[A_{(i)}]+ [H({r'_{03}},\phi_3)].[ B'_{(3)}]=0},
			\end{array}} \right.
			\label{eq:merge-2}
		\end{equation}
		\label{eq:merge}
	\end{subequations}
\end{footnotesize}
Evidently, Eqs. (\ref{eq:merge-1}) and (\ref{eq:merge-2}) have the same form; for scenarios (d) and (f) in Fig. \ref{fig:compleschem}, the above equations produce the same solutions as $[B_{(1)}]=[B'_{(1)}]$ and $[B_{(3)}]=[B'_{(3)}]$. It should be noted that this deduction is established under the assumption that the obstacles have PEC boundaries, which makes the right sides of these equations remain zero during the deduction. For other types of boundary conditions, the analysis becomes more complicated, and more research should be performed.

To show the equivalence between scenarios (d) and (f), the following translation relation \cite{chew1995waves} should be applied to Eq. (\ref{eq:Esab})
\begin{widetext}
	\begin{footnotesize}
		\begin{equation}
			b_{n}^{(1)}{H_n^{(2)}}({k_0}f_3(|\overrightarrow {{r}}-\overrightarrow {{r'_{01}}}|)){e^{in{\phi_{(\overrightarrow {{r}}-\overrightarrow {{r'_{01}}})}}}}=\sum\limits_{m =  - \infty }^\infty b_{n}^{(1)}{J_{m-n}({k_0}f_3(|\overrightarrow {{r'_{01}}}|))H_m^{(2)}({k_0}f_3(|\overrightarrow {{r}}|)){e^{-i(m-n)\phi_1}}{e^{im\phi}}},\;r>r_B.
		\end{equation}
	\end{footnotesize}
\end{widetext}
When $[A_{(3)}] =[J(\eta {r_{01}},\phi_1)].[B_{(1)}]$ in Eq. (\ref{eq:metaMatrix}), Eq. (\ref{eq:Esab}) will be exactly the same as Eq. (\ref{eq:Es1}). In other words, for scenarios (d) and (f), the electric fields in the domain $r> r_B$ ($r_B>r_{2}$) are equivalent.

\textbf{To summarize}, we have discussed the scattering properties of mismatched CM considering penetrated cylindrical PEC obstacles and found that scenarios (c) and (e) (or (d) and (f)) in Fig. \ref{fig:compleschem} will be indistinguishable for the viewer in the domain $r>Max(r_{min},r_2)$ (where $r_{min}=Min(r_B, \eta r_A)$). In fact, the multiple scattering method can also be applied to the analysis of other shapes of obstacles through a multipole expansion, where similar conclusions like those mentioned above can also be drawn. In addition, it should be noted that when $r_B>\eta r_A$ (scenarios (c) and (e)), the equivalence does not rely on the type of boundary for the obstacles; however, when $r_B<\eta r_A$ (scenarios (d) and (f)), the equivalence is derived under the PEC approximation (for obstacles). Other conditions with a different boundary for the obstacles might make the equivalence invalid \cite{2015arXiv150802213Z}.

\section{Numerical verification}
\begin{figure}
	\centering
	\fbox{\includegraphics[width=\linewidth]{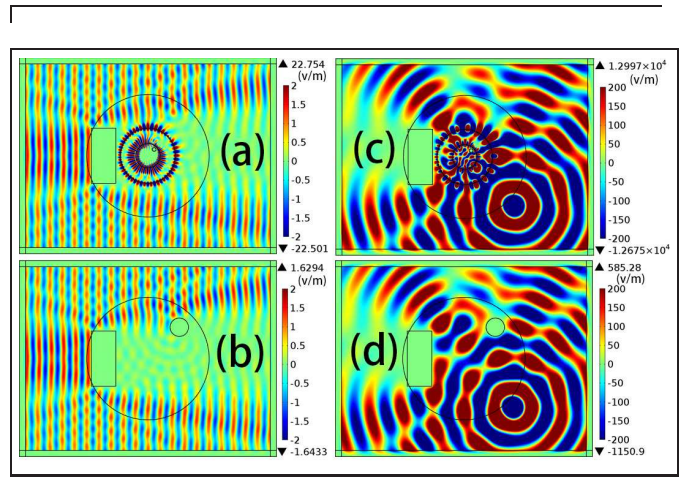}}
	\caption{Numerical verification from COMSOL, for the z component of the electric field. All obstacles have PEC boundaries. Object B (the rectangular object) is located within $r_B<\eta r_A$. (a) For the detection of a TE-polarized plane wave, an amplified image of object A (the circular one inside the ss-shell) could still be detected, as the domain $r>r_B$ in both (a) and (b) possess the same EM fields. Moreover, a point source is applied in (c) and (d).}
	\label{fig:efieldz}
\end{figure}

In addition to the analytical analysis given above, the numerical results from COMSOL shown in Fig. \ref{fig:efieldz} are also provided to verify our conclusion. To distinguish two obstacles, object B is a rectangular cylinder, object A is a circular cylinder, and both have PEC boundaries. From an investigation of the EM field in the domain $r>r_{min}$ (here, $r_{min}=r_B<\eta r_A$), an amplified image of object A can be detected in Fig. \ref{fig:efieldz}(a) and (c). Scenarios (a) and (b) (or (c) and (d)) have the same scattering field for the same EM wave detection (both a plane wave and point source are presented), and further study finds that loss tangents in the NIM up to $0.001$ will not have a significant impact on the results.

\section{Discussion and conclusion}
In conclusion, when arbitrarily situated obstacles occupy the space that used to be canceled by the NIM, mismatched CM are formed. A discussion of such cases is usually ignored or avoided. From our analytical analysis, a clear understanding of the scattering properties of mismatched CM was presented, and numerical results verified the expected effects. More prosaically, we studied a superlens and superscatterer considering penetrated PEC obstacles and found the cancellation ability that exists in ordinary CM was still available in the mismatched case, whereas the ordinary cancellation strategy cannot be applied. Moreover, when obstacles become too close to the NIM, a rigorous analysis showed that the cancellation ability might be established only when a PEC boundary is applied to the obstacles, e.g., for scenario (d) [Fig. \ref{fig:compleschem}], where $r_B<\eta r_A$. The equivalence between scenarios (d) and (f) [Fig. \ref{fig:compleschem}] will rely on the PEC boundary of the obstacles; if other types of boundary conditions are applied, the equivalence might not be valid \cite{2015arXiv150802213Z}. Although most optical devices could be discussed under the PEC approximation, the extension of the scope of the conclusion derived from the PEC-penetrated mismatched CM should be carefully analyzed under the analytical framework provided in this paper. 

On the other hand, the concept of mismatched CM can be applied to the study of the interaction between CM and the penetrated obstacles, which can be encountered in many applications, e.g., CM-based wireless power transfer \cite{superlensWPT,2015arXiv150802213Z}, where the CM could help enhance the transfer efficiency. In fact, the interaction between the emitter, CM, receiver, and even the obstacles in the environment (that may have an impact on the system) can be discussed according to the concept of the mismatched CM. It should be noted, when obstacles are located relatively far away from the NIM, e.g., $r_B>\eta r_A$, the conclusion and equivalence derived above can still be applied for other types of boundary conditions. 

Furthermore, the analytical analysis is not restricted to any specific frequencies, and the conclusion could be applied to studies utilizing a broad range of wavelengths, including applications to CM-modified active cloaking, waveguides, antennas etc. Although the results are obtained for the heuristic 2D case, it is expected to be applicable for more general cases, even for 3D models. Although the numerical results from the FEM (COMSOL) could help illustrate our conclusion, owing to the nonmonotonic transformation, the FEM results might not be reliable and contain large errors in some cases \cite{aznavourian2014morphing,2015arXiv150802213Z}. On the other hand, a strong field exists near the surface of the NIM \cite{yang2008superscatterer}, even for weak EM wave detection; therefore, the transformation optics and the ordinary concept of CM deduced from the long-wavelength limit might not be valid for the mismatched CM. Thus, the analytical framework established in this paper provides an essential tool for future research on mismatched CM.
\section*{Acknowledgments}
\indent This work was sponsored in part by the National Natural Science Foundation of China under Grant No. 51277120 and by the International S\&T Cooperation Program of China (2012DFG01790).
\bibliography{MCcitation}
\end{document}